# Coulomb Blockade Effects in a Topological Insulator Grown on a High-$T_c$ Cuprate Superconductor

Bryan Rachmilowitz[1], He Zhao[1], Zheng Ren[1], Hong Li[1], Konrad H. Thomas[2], John Marangola[1], Shang Gao[1], John Schneeloch[3], Ruidan Zhong[3], Genda Gu[3], Christian Flindt[4] and Ilija Zeljkovic[1].

[1]Department of Physics, Boston College, 140 Commonwealth Ave, Chestnut Hill, Massachusetts 02467

[2]Department of Theoretical Physics, University of Geneva, 1211 Geneva, Switzerland

[3]Brookhaven National Laboratory, Upton, New York 11973, USA

[4]Department of Applied Physics, Aalto University, 00076 Aalto, Finland

**Abstract**

The evidence for proximity-induced superconductivity in heterostructures of topological insulators and high-$T_c$ cuprates has been intensely debated. We use molecular beam epitaxy to grow thin films of topological insulator $Bi_2Te_3$ on a cuprate $Bi_2Sr_2CaCu_2O_{8+x}$, and study the surface of $Bi_2Te_3$ using low-temperature scanning tunneling microscopy and spectroscopy. In few unit-cell thick $Bi_2Te_3$ films, we find a V-shaped gap-like feature at the Fermi energy in dI/dV spectra. By reducing the coverage of $Bi_2Te_3$ films to create nanoscale islands, we discover that this spectral feature dramatically evolves into a much larger hard gap, which can be understood as a Coulomb blockade gap. This conclusion is supported by the evolution of dI/dV spectra with the lateral size of $Bi_2Te_3$ islands, as well as by topographic measurements that show an additional barrier separating $Bi_2Te_3$ and $Bi_2Sr_2CaCu_2O_{8+x}$. We conclude that the prominent gap-like feature in dI/dV spectra in $Bi_2Te_3$ films is not a proximity-induced superconducting gap. Instead, it can be explained by Coulomb blockade effects, which take into account additional resistive and capacitive coupling at the interface. Our experiments provide a fresh insight into the tunneling measurements of complex heterostructures with buried interfaces.

**Introduction:**

When a superconductor (SC) is interfaced with a normal, non-superconducting material, Cooper pairs are able to tunnel across the interface, and the normal material can become superconducting via the superconducting proximity effect (SPE). In the past decade, interest in the SPE has been brought to the forefront by the discovery of new 2D materials predicted to harbor novel electronic phenomena if interfaced with superconductors, such as Majorana modes in topological superconductors [1] and superluminescence in proximitized p-n junctions [2]. A wide range of different SCs have been used as a scaffolding in these efforts, such as $NbSe_2$ [3–7], elemental SCs [8,9], $FeTe_{0.55}Se_{0.45}$ [10,11] and copper-oxide (cuprate) high-$T_c$ SC $Bi_2Sr_2CaCu_2O_{8+x}$ (Bi-2212) [12–17]. While the majority of SCs used have led to well-established platforms for proximity studies, the experiments using cuprates, which exhibit significantly larger superconducting gap ($\Delta_{SC}$) and $T_c$ [18], have given unexpectedly conflicting results [12–17]. On one side, the majority of angle-resolved photoemission spectroscopy (ARPES) measurements reported the absence of an induced gap in the TI band structure, attributed to the very short coherence length of Bi-2212 along the *c*-axis [12], and the mismatch between the Fermi surfaces of the TI and Bi-2212 [12,13]. On the other hand, tunneling measurements have observed a gap in dI/dV spectra in both Bi [17] and $Bi_2Te_3$ [16] grown on Bi-2212, interpreted to arise due to proximity-induced superconductivity in the topological material.

In this work, we provide strong evidence suggesting that Coulomb blockade effects, which arise due to strong electron-electron interactions [19], not proximity induced Cooper pairing, govern the formation of the observed gap at the Fermi level in $Bi_2Te_3$ grown on Bi-2212. This is largely unexpected, because Coulomb gaps have typically been associated with highly-localized quantum dot like structures [20] and systems in high magnetic field [21], not continuous, millimeter-scale heterostructures at zero field such as the ones studied here. However, our experiments reveal that despite continuity of our thicker films (up to ~4 nm thickness in Fig. S7) over macroscopic length scales, inevitable domain formation in this system and an imperfect interface can lead to the same mesoscopic effect in tunneling measurements.

**Results:**

$Bi_2Te_3$ films are grown on UHV-cleaved, optimally-doped Bi-2212 ($T_c$~91 K) using molecular-beam epitaxy (MBE), following a recipe similar to previous related work [12,13] (Supplementary Information). As the proximity-induced superconducting gap at the exposed "bare" surface of the normal material is expected to decrease with its thickness [22], we explore a range of $Bi_2Te_3$ thicknesses, from ~10 quintuple layers (QLs) down to a partial coverage of a single QL film. There are two main challenges in growing $Bi_2Te_3$ on superconducting Bi-2212. The first one is incompatibility of the in-plane atomic structures of the two materials. Bi-2212 cleaves between two BiO planes to reveal a square lattice of Bi atoms observed in STM topographs [Fig. 1(b)], which is in contrast to $Bi_2Te_3$ that has a hexagonal lattice structure [Fig. 1(c)]. This structural mismatch leads to the formation of two types of $Bi_2Te_3$ domains that are rotated 30 degrees with respect to each other [Fig. 1(e), Supplementary Fig. S1] [11]. Despite the domain formation, we are still able to routinely locate large (at the order of ~50 nm squared) single-domain areas using STM in thicker films. Second crucial issue is that the electronic properties of the Bi-2212 surface can change upon heating [23]. At ideal $Bi_2Te_3$ growth temperatures of ~350 °C [24], interstitial oxygen dopants escape from the topmost surface layers of Bi-2212, leading to an effective lowering of the hole density and degradation of superconducting properties [23]. To mitigate this issue, in this work we use much lower $Bi_2Te_3$ growth temperatures below 250 °C (Supplementary Information). Our ~0.1 QL film leads to the formation of nano-islands allowing us to verify the superconducting properties of our Bi-2212 substrate after it has been exposed to our growth conditions [Supplementary Fig. S2].

To characterize the large-scale electronic structure of our $Bi_2Te_3$ films, we use quasiparticle interference (QPI) imaging [25,26], rooted in elastic scattering and interference of electrons on the surface of a material, which can be seen as waves in STM dI/dV maps. We observe the characteristic scattering pattern and energy dispersion, which is qualitatively similar to that reported on cleaved bulk $Bi_2Te_3$ single crystals [26,27], as well as $Bi_2Te_3$ films [11,28] [Supplementary Fig S3, S4]. Next, we look for any evidence of induced superconductivity at the surface of $Bi_2Te_3$ films, starting with thinner films. dI/dV spectra acquired on 1 QL and 2 QL $Bi_2Te_3$ films show a small suppression in the local density of states within ± ~5-10 meV around the Fermi level [Fig. 2(b,e)]. To test if the surface state is gapped within the energy range of the gap-like feature, we acquire dI/dV maps at small bias near the Fermi level energy, well within the observed gap feature [Fig. 2(a,d)]. Given that we observe a prominent 6-fold QPI signal due to hexagonal warping of the Dirac dispersion starting at ~200 meV below the Fermi level to several hundred meV above the Fermi level (Fig. S3), if the surface state is indeed fully gapped within the narrow energy range across the Fermi level, we would expect to observe a strong suppression of the QPI signal. However, the FTs of the dI/dV maps in both films show a noticeable QPI signal [insets in Fig. 2(b,e)], which allows us to place an upper bound on the magnitude of the induced gap in the $Bi_2Te_3$ surface state, if any, to ~2-3 meV, set by the finite temperature thermal broadening and lock-in excitation used in our experiments. In comparison to the average dI/dV spectrum of optimally-doped Bi-2212 that shows ~40 meV magnitude and prominent coherence peaks [Supplementary Fig. S2(c)], the gap-like feature observed on $Bi_2Te_3$ is shallow, often asymmetric and does not show clear coherence peaks [Fig. 2(c,f)]. Previous experiments have attributed qualitatively similar gap-like feature to an induced superconducting gap in the normal material in proximity to cuprates [16,17].

Importantly, films of reduced coverage below 1 QL enable us to investigate bonding of the two materials at the interface. We analyze STM topographs in more detail to extract their topographic height with respect to the substrate [Fig. 3(a)]. Interestingly, we find the apparent height of all 1 QL $Bi_2Te_3$ islands to be ~1.2 nm with respect to Bi-2212 [Fig. 3(d)], ~20% taller than the expected ~1 nm height for a single QL of $Bi_2Te_3$. The same result is confirmed in a film with a nearly complete 1 QL coverage, where the island height is also found to be ~20% larger than expected [Fig. 3(b,e)]. To demonstrate that our STM scanner has been properly calibrated, we plot the topographic height profile across a step in the 2 QL thick film with respect to the $Bi_2Te_3$ layer below, which shows the expected step height of ~1 nm [Fig. 3(c,f)]. As STM topographs contain both electronic and structural information, to provide evidence that this height difference is of structural origin, and not purely due to the variation in electronic density of states, we provide the following pieces of evidence. First, the same island height is extracted from topographs acquired at both positive and negative STM bias [Supplementary Fig. S5]. Second, 1 QL $Bi_2Te_3$ nano-islands grown on a superconductor Fe(Te,Se) and on an insulator $SrTiO_3$(001), two materials that are electronically very different, yield a ~1 nm step height in both cases [Supplementary Fig. S6]. To explain the larger height of the first $Bi_2Te_3$ layer with respect to the Bi-2212 substrate, we postulate that there may be an inter-growth layer forming at the interface [Fig. 3(g-i)]. This intergrowth layer could be composed of dilute amounts of excess Bi or Te, possibly the same as the small clusters that we see in topographs of exposed regions of Bi-2212, which are also approximately equal to the barrier height (Fig. S2). This has occasionally been observed at the interface of other van der Waals heterostructures [29].

We examine the consequences of this imperfect interface in the context of TI/SC heterostructures. A crucial insight comes from investigating the electronic properties of $Bi_2Te_3$ nano-islands themselves [Fig. 4(a,b)]. The average dI/dV spectrum acquired on a small island with ~10 nm diameter [Fig. 4(a)] shows a markedly different shape compared to that observed on a much larger flat terrace of 1 QL $Bi_2Te_3$ film [Fig. 2]. We observe a hard gap spanning the Fermi level with ~150 meV magnitude, much larger than $2\Delta_{SC}$ of Bi-2212. Interestingly, the gap displays a pronounced asymmetry in energy, which as we will show can be attributed to a residual charge on the island (Fig. S10). Due to this asymmetry, and the fact that a gap as large as several hundred meV can be seen on other islands, we rule out the superconducting origin of the gap.

Given the additional barrier at the interface we have discovered and the finite size of the system, a natural explanation for the gap could be understood in terms of a Coulomb blockade (CB) gap [30], which arises due to single electrons exchanging energy with the environment as they tunnel through a barrier [31,32]. The CB effect has been widely reported in tunneling measurements of finite-size heterostructures when there is an extra barrier at the interface of the two materials sandwiched together [33–36]. It can be modeled by a double tunnel junction, one being the tip-sample junction and the other one being the sample-substrate junction, each consisting of a capacitor and a resistor connected in parallel [Fig. 4(d)]. The overall size of the CB gap ($\Delta_{CB}$) is roughly inversely proportional to the capacitance between the film and the substrate ($C_2$) [33,34]. The shape of the gap will change depending on the resistive component at the film-substrate junction ($R_2$), where large $R_2$ would lead to a sharp-cutoff in conductance at the gap edge, while smaller $R_2$ would lead to a gradual suppression of conductance approaching zero energy [33,34].

In small $Bi_2Te_3$ islands, due to small $C_2$ and large $R_2$, we observe a sharp CB gap [Fig. 4(a)], with zero dI/dV conductance within the gap. As the size of the island increases ($C_2$ should become larger and $R_2$ smaller), $\Delta_{CB}$ is expected to evolve into a smaller V-shaped gap [33,34]. This is exactly what we observe in our data [Fig. 4(b)]. As the islands grow even larger and start to connect, the V-shape gap-feature becomes progressively more subdued [Fig. 4(c)], with finite conductance near the Fermi level where we are still able to observe a QPI signal. Similarly to Ref. [33,34], we use the dynamical Coulomb blockade P(E) theory to fit the overall spectral shape, which show a good agreement for a reasonable set of $C_2$ and $R_2$ parameters (Fig. S11, Fig. S12, Supplementary Information 4). We note that $Bi_2Te_3$ films inevitably contain domains due to structural mismatch between $Bi_2Te_3$ and Bi-2212, observed in reflection high-energy electron diffraction (RHEED) images [Supplementary Fig. S1(b)] and in real space [Supplementary Fig. S1(a)], which can lead to the finite size effects as seen in STM/S measurements, even as the islands merge together in one or more QL thick films. On our thicker films, the gap-like feature gets suppressed and the zero-bias conductance increases [Fig. 5]. Specifically, while the gap-like feature is prominent in 1, 2 and 4 QL films, we only see a small suppression at zero energy in ~9-10 QL films, which nearly completely disappears in the thickest ~11 QL region [Fig. 5]. This can be understood as a consequence of the decrease in resistivity. We also postulate that domains, on average, may become larger in thicker films due to strain relaxation as the films grow thicker. Therefore, the CB effect will still be present, but weaker in thicker films (Fig. S12), and the gap-like feature observed in our measurements is most likely a simple consequence of this phenomenon, not proximity-induced Cooper pairing.

**Discussion and outlook:**

Our measurements resolve the outstanding controversy between ARPES and STM measurements of MBE-grown heterostructures of topological materials and cuprates, by shedding light on an overlooked aspect of underlying physics in these systems rooted in the CB effects. Although CB should in principle be negligible in continuous millimeter-scale heterostructures, we find that tunneling measurements of $Bi_2Te_3$ films can still exhibit a CB gap due to domain formation during growth and the additional barrier at the interface between $Bi_2Te_3$ and Bi-2212. The gap is as large as a few hundred meV in the smallest $Bi_2Te_3$ islands, but as the size of the islands increases, this leads to the mitigation of the CB effect, and a much smaller V-shaped gap in the nearly continuous (but with inevitable domains) $Bi_2Te_3$ films of one or more QL thickness. We note that the complex structure of additional peaks in dI/dV spectra outside the gap [Fig. 4(a)] may be due to quantized bound states within the islands [Supplementary Fig. S8], similarly to what had been observed in quantum dots [37]. Another explanation for these peaks may be due to the capacitive coupling between the tip and the island [36], which could be explored in future experiments and analysis. However, a quantitative understanding of each peak position is beyond the scope of this paper. Our work cautions the interpretation of gap-like features in tunneling measurements of complex heterostructures with inaccessible, buried interfaces.

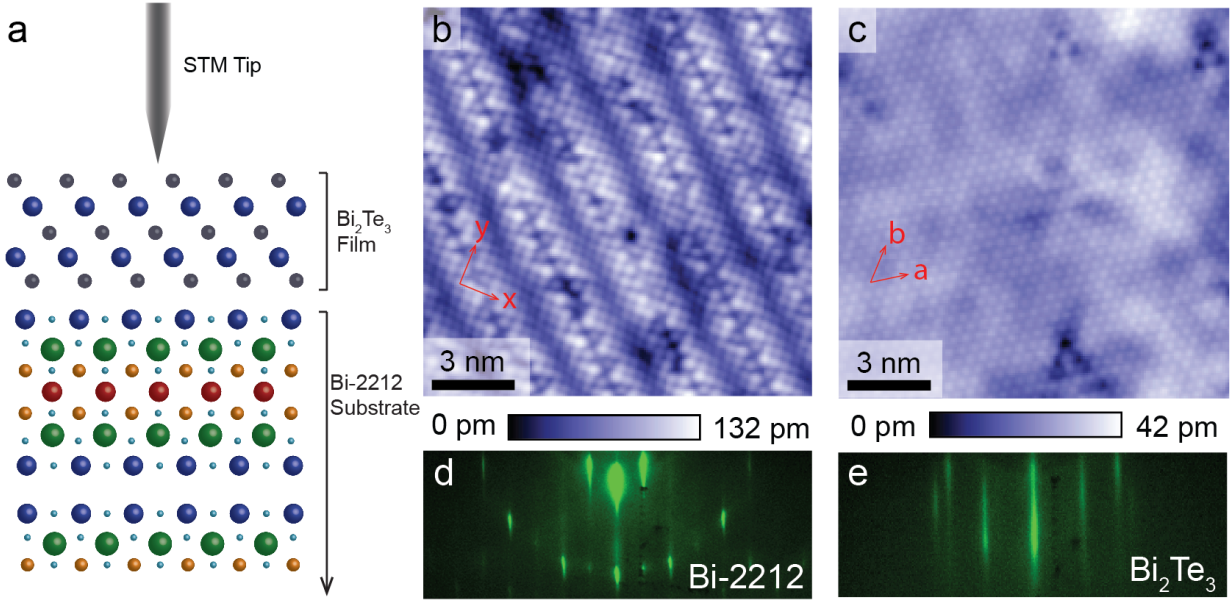

**Figure 1: The schematic of the heterostructure, characteristic topographs and diffraction patterns.** (a) Schematic of the $Bi_2Te_3/Bi_2Sr_2CaCu_2O_{8+x}$ heterostructure. STM topograph showing (b) $Bi_2Sr_2CaCu_2O_{8+x}$ (Bi-2212) substrate before deposition, (c) $Bi_2Te_3$ film after deposition. Reflection high-energy electron diffraction (RHEED) images acquired (d) before and (e) after the $Bi_2Te_3$, growth showing characteristic Bi-2212 and $Bi_2Te_3$ features, respectively. STM setup condition: (b) $I_{set}$ = 10 pA, $V_{sample}$ = 100 mV; (c) $I_{set}$ = 50 pA, $V_{sample}$ = -50 mV.

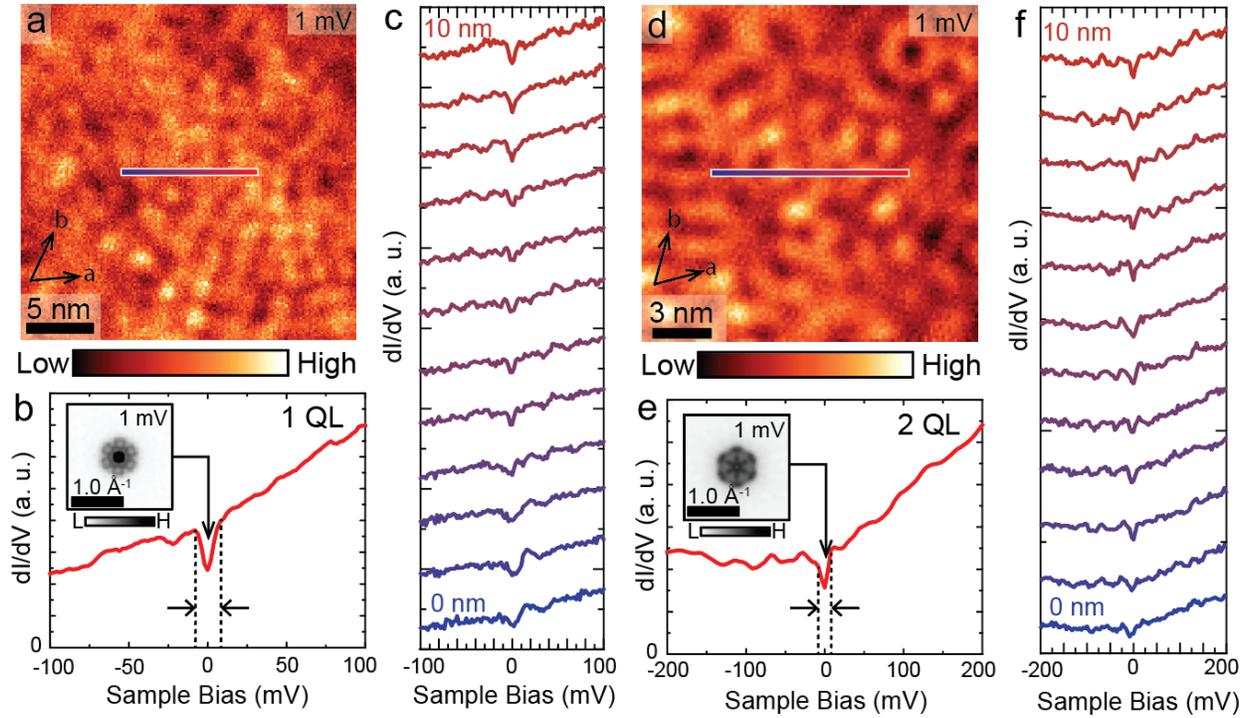

**Figure 2: In-gap quasiparticle interference signature and spatial variations in differential conductance.** (a,d) dI/dV map taken on (a) 1 QL and (d) 2 QL thick $Bi_2Te_3$ films showing distinct QPI signature at a 1 mV bias. (b,e) Average dI/dV spectra obtained on (b) 1 QL and (e) 2 QL films showing a gap like feature at the Fermi energy in the density of states denoted by dashed lines and arrows. Inserts in (b) and (e) are the six-fold symmetrized Fourier transforms of (a) and (d), respectively. (c,f) 10 nm line cut of dI/dV spectra on (c) 1 QL and (f) 2 QL films offset for clarity showing the variation of the gap-like feature. Colored lines in (a,d) denote the positions of the linecuts in (c,f). STM setup conditions: (a) $I_{set}$ = 3 pA, $V_{sample}$ = 1 mV, $V_{exc}$ = 1 mV (zero-to-peak); (b,c) $I_{set}$ = 40 pA, $V_{sample}$ = 100 mV, $V_{exc}$ = 4 mV; (d-f) $I_{set}$ = 100 pA, $V_{sample}$ = 50 mV, $V_{exc}$ = 2 mV.

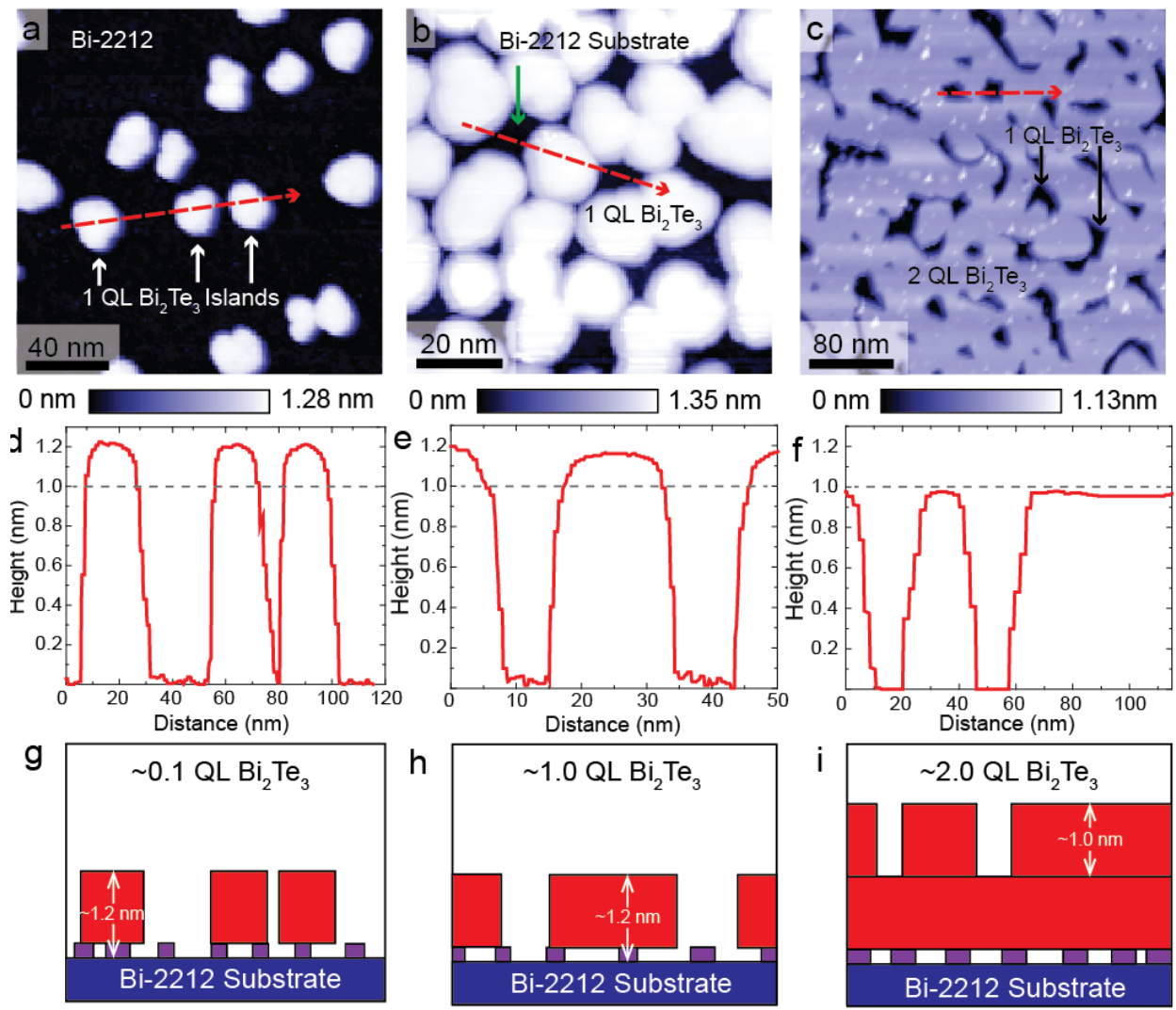

**Figure 3: Topographic heights of Bi₂Te₃ thin films.** (a-c) STM topographs showing (a) Bi$_2$Te$_3$ nano-islands on Bi-2212 with ~10% coverage, (b) Bi$_2$Te$_3$ nano-islands on Bi-2212 with ~90% coverage and (c) ~2 QL Bi$_2$Te$_3$ film. (d-f) Topographic profiles from (a-c) along the lines indicated by red dashed arrows in (a-c). (g-i) Schematic depiction of the cross-section of the interface, consistent with the observed step heights in (d-f). STM setup condition: (a) I$_{set}$ = 10 pA, V$_{sample}$ = -500 mV; (b) I$_{set}$ = 20 pA, V$_{sample}$ = 2 V; (c) I$_{set}$ = 10 pA, V$_{sample}$ = 500 mV.

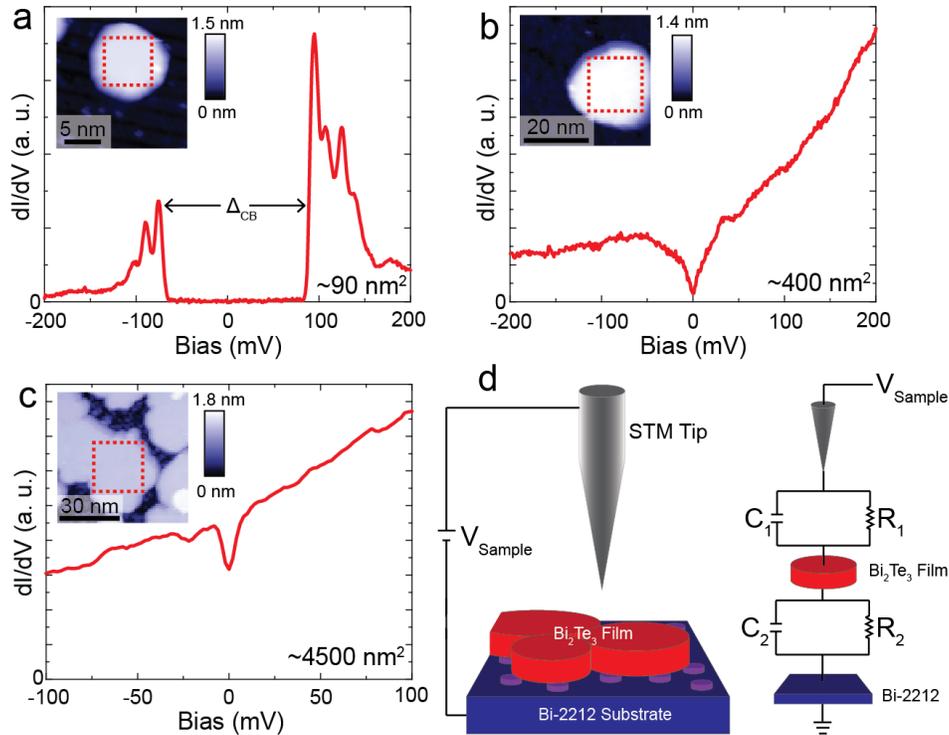

**Figure 4: Evolution of the gap-like feature in the Bi$_2$Te$_3$ single layer.** (a-c) Average dI/dV spectra acquired over the regions denoted by the red dashed squares in the inset topographs. Average dI/dV measured on the surface of (a) ~90 nm$^2$ island, (b) ~400 nm$^2$ island and (c) ~4500 nm$^2$ region of Bi$_2$Te$_3$ nano-island. (d) Schematic of double tunnel junction, where $C_1$ and $C_2$ are the capacitance of tip to film and film to substrate junctions, and $R_1$ and $R_2$ are their respective resistances. STM setup conditions: (a) $I_{set}$ = 50 pA, $V_{sample}$ = 200 mV, $V_{exc}$ = 2 mV, Inset topography: $I_{set}$ = 10 pA, $V_{sample}$ = 1000 mV; (b) $I_{set}$ = 50 pA, $V_{sample}$ = 200 mV, $V_{exc}$ = 2 mV, Inset topography: $I_{set}$ = 10 pA, $V_{sample}$ = -500 mV; (c) $I_{set}$ = 40 pA, $V_{sample}$ = 100 mV, $V_{exc}$ = 4 mV, Inset topography: $I_{set}$ = 5 pA, $V_{sample}$ = 600 mV.

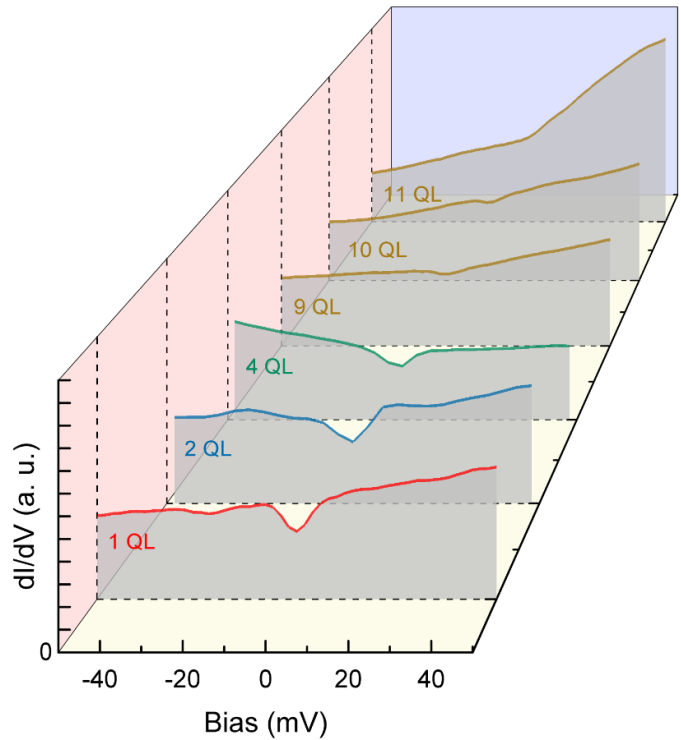

**Figure 5: Thickness evolution of the gap-like feature.** Average dI/dV spectra taken over $Bi_2Te_3$ of various thickness on four different samples: 1 QL, 2 QL and 4 QL films were all different samples, while ~9, ~10 and ~11 QL data was acquired on consecutive terraces of the same sample with ~10 QL nominal thickness. All spectra were normalized by integral of dI/dV from 0 to -50 mV. The difference in the shape of the background conductance of the 4 QL spectrum is likely an artifact of the STM tip with slightly anisotropic density of states. STM setup conditions: (1 QL) $I_{set}$ = 40 pA, $V_{sample}$ = 100 mV, $V_{exc}$ = 4 mV; (2 QL) $I_{set}$ = 100 pA, $V_{sample}$ = 50 mV, $V_{exc}$ = 2 mV; (4 QL) $I_{set}$ = 200 pA, $V_{sample}$ = 200 mV, $V_{exc}$ = 2 mV; (9, 10, 11 QL) $I_{set}$ = 100 pA, $V_{sample}$ = 50 mV, $V_{exc}$ = 2 mV.

**Data Availability**

The data supporting the findings of this study are available upon request from the corresponding author.

**Author Contributions**

MBE growth was performed by B.R. and Z.R. STM experiments were carried out by H.Z, H.L. and S.G. B.R, H.Z. and H.L. analyzed the STM data with the guidance from I.Z. J.S, R.Z. and G.G grew single crystals of Bi-2212. I.Z and B.R wrote the manuscript with input from all the authors. K.H.T. and C.F. provided theory for the P(E)-calculations of the differential conductance. B.R. and J.M. compared theoretical fits with exerimental data. I.Z. supervised the project.

**Competing Interests**

The Authors declare no Competing Financial or Non-Financial Interests.

**Code availability**

The computer code used for data analysis is available upon request from the corresponding author.

**Acknowledgements**

I.Z. gratefully acknowledges the support from Army Research Office Grant Number W911NF-17-1-0399. The work in Brookhaven was supported by the Office of Science, US Department of Energy under Contract No. DE-SC0012704. J.S. and R.D.Z. were supported by the Center for Emergent Superconductivity, an Energy Frontier Research Center funded by the U.S. Department of Energy, Office of Science.